\documentclass[conference]{IEEEtran}
\IEEEoverridecommandlockouts
\usepackage{color}
\usepackage{cite}
\usepackage{enumitem}
\usepackage{amsmath}
\usepackage{url}

\usepackage[binary-units = true, range-phrase=--, per-mode=symbol]{siunitx}
\DeclareSIUnit{\decibelm}{dBm}
\DeclareSIUnit{\Joule}{Joule}

\newcommand{\fakeparagraph}[1]{\vspace{.5mm}\noindent\textbf{#1.}}
\newcommand{\fakepar}[1]{\fakeparagraph{#1}}

\ifCLASSINFOpdf
   \usepackage[pdftex]{graphicx}
\else
\fi

\begin{document}

\title{Over-the-Air Time Synchronization for URLLC: Requirements, Challenges and Possible Enablers}

\author{\IEEEauthorblockN{Aamir Mahmood\IEEEauthorrefmark{1}, Muhammad Ikram Ashraf\IEEEauthorrefmark{2},
Mikael Gidlund\IEEEauthorrefmark{1} and
Johan Torsner\IEEEauthorrefmark{2}}
\IEEEauthorblockA{\IEEEauthorrefmark{1}Department of Information Systems and Technology,
Mid Sweden University}
\IEEEauthorblockA{\IEEEauthorrefmark{2}Ericsson Research, Finland}
Email: \IEEEauthorrefmark{1}firstname.lastname@miun.se, \IEEEauthorrefmark{2}\{ikram.ashraf, johan.torsner\}@ericsson.com
}


\maketitle

\begin{abstract}


Ultra-reliable and low-latency communications (URLLC) is an emerging feature 
in 5G and beyond wireless systems, which is introduced to support stringent 
latency and reliability requirements of mission-critical industrial 
applications. In many potential applications, multiple sensors/actuators 
collaborate and require isochronous operation with strict and bounded jitter, 
e.g., \SI{1}{\micro\second}. To this end, network time synchronization 
becomes crucial for real-time and isochronous communication between a 
controller and the sensors/actuators. In this paper, we look at different 
applications in factory automation and smart grids to reveal the requirements 
of device-level time synchronization 
and the challenges in extending the high-granularity timing information to the 
devices. Also, we identify the potential over-the-air synchronization 
mechanisms in 5G radio interface, and discuss the needed enhancements to meet the jitter 
constraints of time-sensitive URLLC applications.
\end{abstract}

\IEEEpeerreviewmaketitle

\section{Introduction}



In 5G and beyond wireless systems, ultra-reliable and low-latency communications (URLLC) 
feature is focused on time-sensitive applications originating for vertical 
industries such as industrial automation, smart grids, tactile internet, 
automotive and more. There can be many use cases within a single industry 
while each use case presents a different set of requirements and challenges. For 
instance in industrial automation, factory automation is one of the most 
challenging use case for URLLC that requires deterministic communication with 
bounded reliability and latency. In addition, factory automation often 
entails real-time interactions among multiple entities, and ultra-tight 
synchronization of the entities with a common time reference is need to complete manufacturing. As there are many existing industrial 
wired and wireless systems~\cite{frotzscher2014}, 5G radio access requires a new time 
synchronization service for enabling URLLC in heterogeneous industrial setups.



\begin{figure}[!t]
	\centering
		\includegraphics[width=1.0\linewidth]{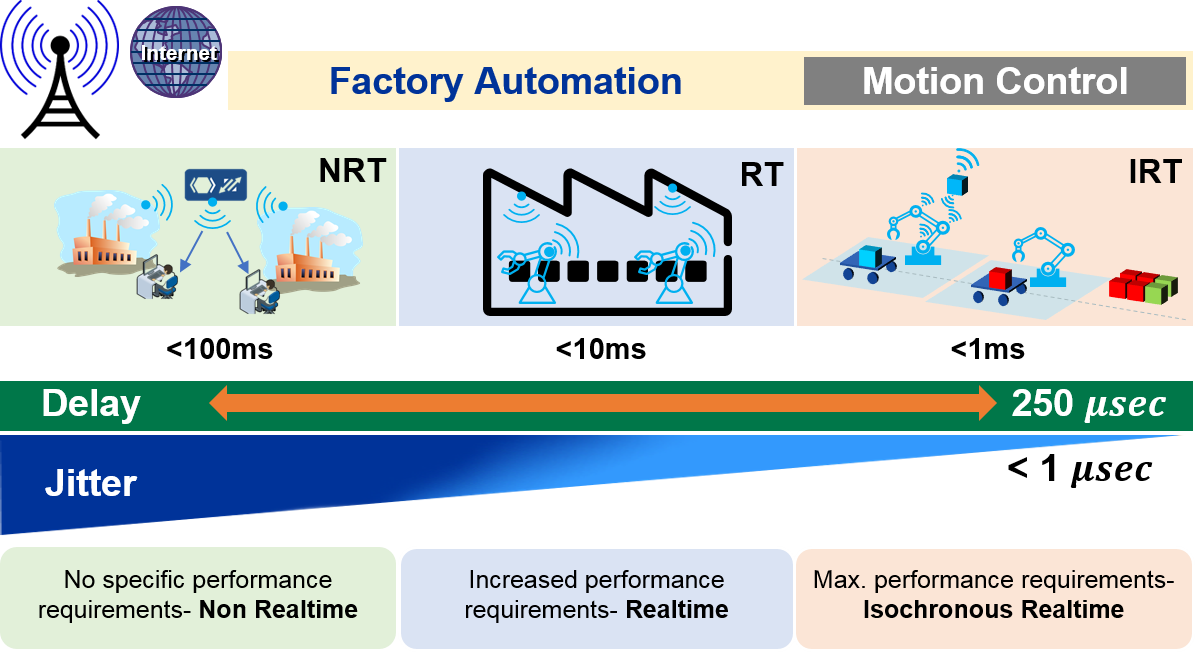}
		\vspace{-15pt}
	\caption{Performance requirements of an industrial communication system.}
	\label{fig:Req}
	\vspace{-10pt}	
\end{figure}

In factory automation, when classifying applications regarding their 
timeliness and reliability requirements, there can be three main application 
classes; i) non real-time (NRT) or soft real-time, ii) hard real-time (RT), and 
iii) isochronous real-time (IRT) as illustrated in Fig.~\ref{fig:Req}. Most of 
the applications in process automation belong to NRT or soft real-time class. 
Whereas the discrete manufacturing applications, which rely on robotics and 
belt conveyers for assembly, picking, welding and palletizing, execute tasks 
in a timely and sequential manner \cite{aakerberg2011future}. These 
jobs involve tightly synchronized real-time cooperation among multiple robots 
and the production line. To generalize, discrete manufacturing applications are a 
part of hard RT or IRT class; that is, given deadlines must be met strictly 
or deadlines must be satisfied along with the constraints on 
jitter. In this regard, IEEE 802.1 time-sensitive networking (TSN) standards specify 
strict performance requirements: 
\SI{1}{\milli\second} cycle time, \SI{99.999}{\percent} reliability and 
\SI{1}{\micro\second} jitter i.e., allowed variations in delay. 

To satisfy 
the needed \textit{determinism} and 
\textit{synchronism} in industrial automation, 5G URLLC requires innovative 
solutions. Several radio access 
solutions are under consideration to meet the latency and reliability 
requirements (e.g., see \cite{URLLCtail, SachURLLC}). However, RT and IRT 
communication with tight jitter constraint requires accurate time 
instants on 
a common time base at the device level. In a factory floor, the devices might 
belong to different base stations (BS) or even to different domains in case 
of coexisting industrial networks. To get a common time reference 
for the devices would require over-the-air (OTA) synchronization mechanisms 
beyond timing advance-based frame alignment 
between UEs and BS. While, OTA synchronization procedures for LTE-TDD and 
radio coordination in small cells are limited to synchronization of BSs 
only \cite{SyncSmallCell}. To 
enable time synchronization for the devices, a recent LTE URLLC work item includes that the  
synchronization shall be standardized for LTE Rel-15~\cite{3gpp.WID}.



%


 %
%
%


In this article, we highlight the drivers and challenges of ultra-tight time 
synchronization in factory automation and smart grids. We identify the 
opportunities in the LTE air interface to enable OTA synchronization 
solutions. In particular, we summarize how these signaling 
parameters can be enhanced and grouped together to extend high-granularity 
timing information to the devices. 

\section{Preliminaries}
\subsection{Oscillator, Clock, Synchronization, Accuracy, Jitter etc.}

\begin{figure}[!t] 	
	\centering 		
	\includegraphics[width=0.85\linewidth]{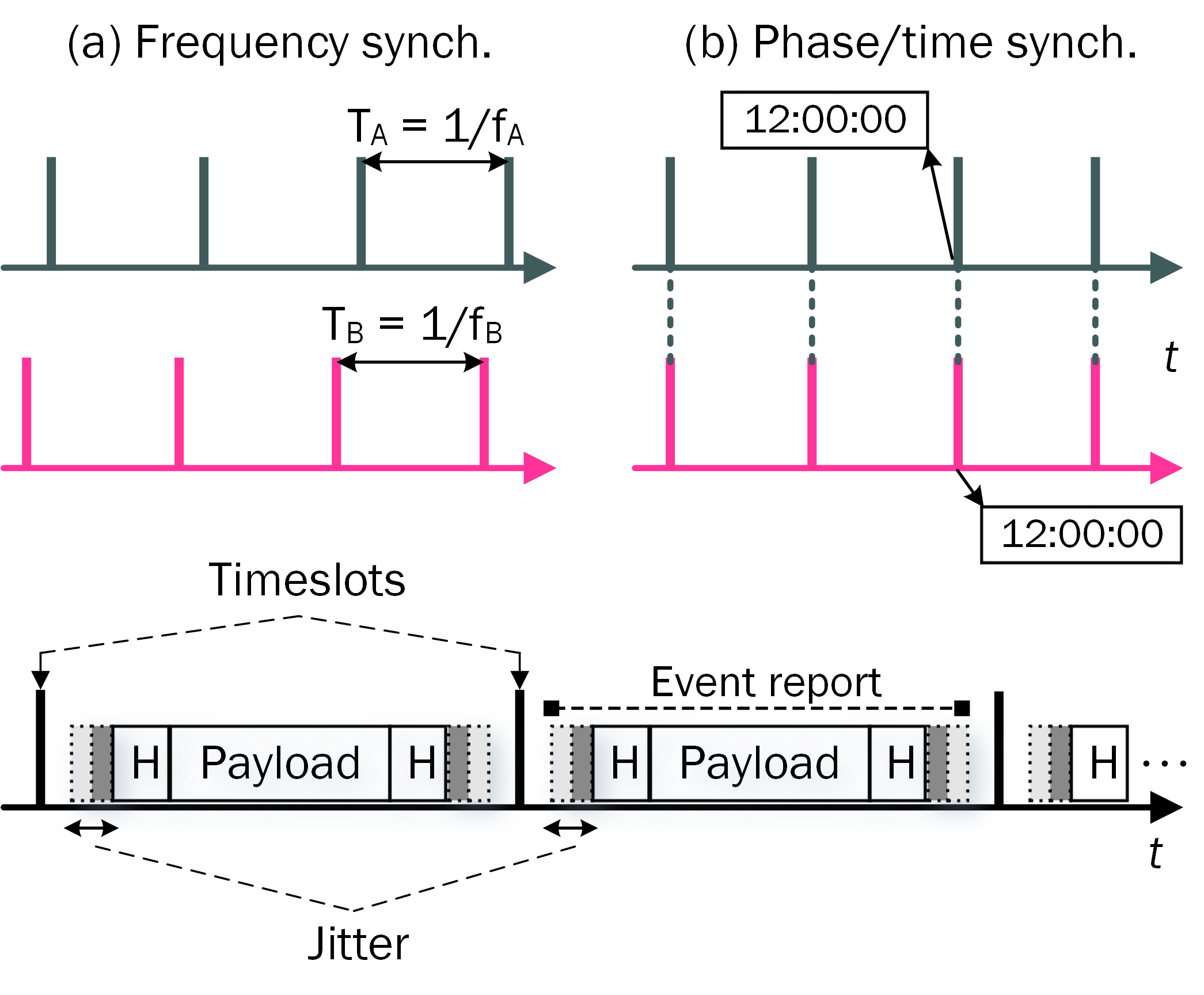}
	\vspace{-12pt}
 	\caption{A sketch (top) showing two systems A and B maintaining: (a) frequency synchronization ($f_A = f_B$), (b) phase and time synchronization. The bottom sketch shows jitter in packet reception at the controller.} 	
	\label{fig:synctypes} 
	\vspace{-10pt}
\end{figure}

\fakepar{Synchronization types} ITU-T G.8260 defines three types of synchronization: 
\begin{itemize}[leftmargin=*]
	\item \textit{Frequency}: two systems are frequency synchronized when their 
significant instants occur at the same rate.
	\item \textit{Phase}: in phase synchronized systems, the rising edges occur 
at the same time, e.g., the point in time when the time slot of a frame is to 
be generated.
	\item \textit{Time}: time synchronization is the distribution of an 
absolute time reference to a set of real-time clocks. The synchronized clocks 
have a common epoch timescale. Note that distribution of 
time synchronization is a way of achieving phase synchronization.
\end{itemize}
The synchronization types are illustrated in Fig.~\ref{fig:synctypes}. As we focus on time-aware URLLC applications, the further discussion is 
confined to time synchronization.

\renewcommand{\fakeparagraph}[1]{\vspace{.5mm}\noindent\textbf{#1}}
\renewcommand{\fakepar}[1]{\fakeparagraph{#1}}

\fakepar{What causes time de-synchronization?} A device maintains the 
sense of time--a clock--by counting the pulses of an internal 
crystal oscillator \cite{RecursiveClockSkewRB}. But, there is an inherent 
inaccuracy in frequency (causing clock skew) and phase of the 
crystal oscillator. The inaccuracy is influenced by the operating conditions and 
aging (resulting in clock drift). As a result, the devices deviate from 
a reference clock after a synchronization epoch~\cite{muSync}.

\renewcommand{\fakeparagraph}[1]{\vspace{.5mm}\noindent\textbf{#1.}}
\renewcommand{\fakepar}[1]{\fakeparagraph{#1}}

\fakepar{Jitter} Implies the packet delay variations from defined ideal 
position in time (Fig.~\ref{fig:synctypes}). Many closed-loop control 
applications are intolerant to such delay variations.
  
\subsection{Time Synchronization in Industrial Networks}

\begin{figure}[!t] 	
	\centering 		
	\includegraphics[width=1\linewidth]{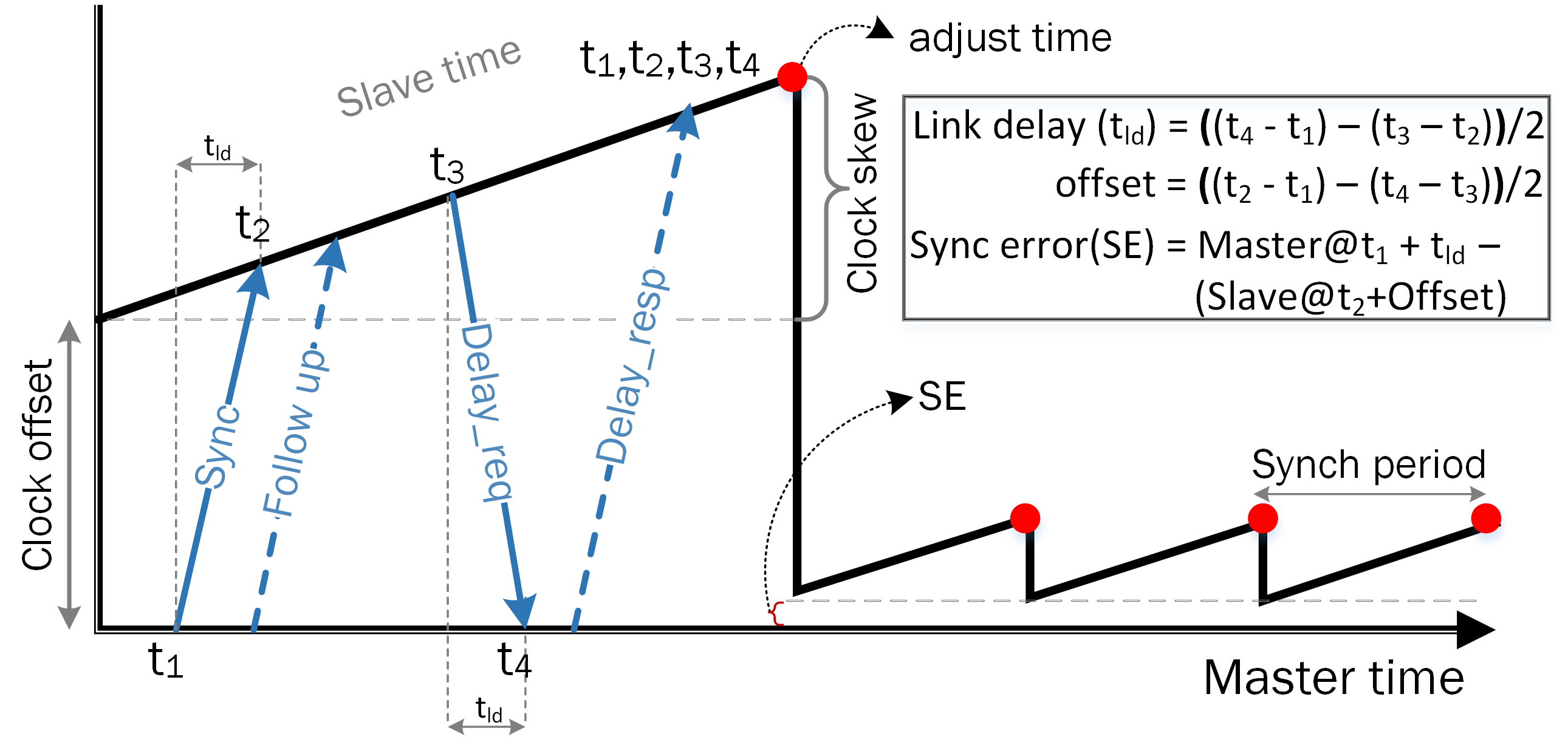}
	\vspace{-15pt}
 	\caption{Time synchronization procedure in precision time protocol (PTP). Even after synchronization, there can be synchronization error due to inaccuracy in timestamping and asymmetric link delays. Also, the devices need to be synchronized periodically while the synchronization period depends on the clock stability and required accuracy.} 	
	\label{fig:ptp} 
	\vspace{-10pt}
\end{figure}

Networked distributed measurement and control systems (e.g., bus systems and 
factory automation setups) are reliant on IEEE 1588 precision time protocol 
(PTP) \cite{IEEE_Std_1588_2008} for real-time and isochronous transmissions. 
IEEE 1588 is a master-slave protocol designed to synchronize real-time clocks 
in the order of sub-microseconds in packet-based networks. Fig.\ref{fig:ptp} 
shows the synchronization procedure, which is based on the 
timestamping of the exchanged signaling messages to find time offset of a 
device from the master clock.

Ethernet-based automation networks, PROFINET and IEEE 802.1 
TSN, use IEEE 1588 
variants (PTP profile) to ensure determinism. The 802.1 TSN task group~\cite{TSN2} is 
developing standards to support time-sensitive applications, for industries 
like factory automation and automotive, over IEEE 802 
networks. TSN contains a series of standards related to ultra reliability, 
low latency and resource management aspects while TSN 802.1AS is the 
standard for transport of precise timing and synchronization
\cite{IEEE_Std_8021AS_2011}. Note that Wi-Fi \textit{TimeSync} 
also includes an extension of 802.1AS~\cite{wi-fitimesync}. 

\section{Drivers for Ultra-Tight Synchronization and Requirements}
Two important use cases of URLLC are industrial factory automation and smart 
grids \cite{3gpp.22.804}. These use cases have strict jitter requirements 
among devices as we discuss next.

\subsection{Industrial Factory Automation or Automation Control}

In factory automation, closed-loop control applications--robot manufacturing, 
round table production, machine tools, packaging and 
printing etc.--are the main URLLC targets. In these applications, the need 
for 
ultra-tight synchronization is driven by \textit{real-time and time-slotted 
communication}, and \textit{isochronous task execution}.

\fakepar{Real-time communication}  
A typical closed-loop control cycle consists of a downlink transaction to a 
set of sensors periodically, which is followed by uplink responses by the 
sensors to the controller. The control events are executed within a 
\textit{cycle time} and may occur isochronously. Cycle time is the 
time from the transmission of a command by the controller to the reception of 
its response from the devices. TSN~\cite{IEEE_Std_8021AS_2011} specifies 
stringent end-to-end latency and reliability constraints on each transaction 
i.e., it must be completed within a \SI{1}{\milli\second} cycle time with 
\SI{99.999}{\percent} reliability. Furthermore, a jitter constraint of 
\SI{1}{\micro\second} is imposed on the delivery of responses which requires 
better than \SI{+-500}{\nano\second} synchronization accuracy among devices. 
These requirements are also endorsed by 3GPP~\cite{3gpp.22.804}. Any 
violation of latency and jitter in control commands can damage the production 
line and cause the safety issues.

\fakepar{Multi-robot cooperation--isochronous real-time} In motion control 
applications--mobilizing a fleet of tractors, symmetrical welding and 
polishing in the automobile production line--a group of robots 
collaborate to execute meticulously sequenced functions. A critical 
requirement to carry out these cooperative actions is synchronous task 
execution, which requires accurate time base across the collaborating 
entities. Therefore, when a controller sends a command to the robots to act 
at a specific time instant (isochronously), the robots should act/respond in 
less than \SI{1}{\micro\second}. A lag in action may cause damages or inefficient production. 

\fakepar{Time-slotted communication} To transport packets with bounded delays 
in RT traffic, time-slotted communication is an effective mechanism. Existing 
industrial wired and wireless networks implement a variant of time-slotted 
communication. It requires perfect time synchronization as any timing error 
leads to the overlap of time slots among the devices, which disrupts the 
communication reliability.

In monitoring applications, the time information must be embedded into the 
sensory data for operations such as data fusion. Thus, the collaborating 
sensors must be synchronized.    

\subsection{Smart Grids}
\label{subsec:SmartGrids}

The traditional power grid is rapidly evolving to a smart grid through the 
automation of control and monitoring functions. To support this 
evolution, a cost-effective wireless communication infrastructure with 
wide-area coverage and high performance is needed. 5G offers the needed coverage 
while the performance requirements are application dependent, hence, needing 
further study. Three main application areas are \textit{fault protection}, 
\textit{control}, and \textit{monitoring and diagnostics} \cite{ABB5GSG}. 

\fakepar{Fault protection} High-speed communication of 
measurements between the two points of a transmission line can detect a fault 
using a line differential protection solution \cite{ABB5GSG}. In line differential 
protection, two relay devices periodically sample the electric current in a section of distribution or transmission 
line, and exchange this information with each other. In case of 
a fault, the relays differ in measurements and can trigger a trip command to 
the breaker. Fault protection has the most stringent performance requirement: 
reliability \SI{> 99.99}{\percent} and latency \SI{< 10}{\milli\second}. 
In addition, the protection algorithm is effective if the 
relays are tightly synchronized (i.e., \SI{< 20}{\micro\second}).     

\fakepar{Control and grid automation} In evolving power grids, with high 
penetration of renewable resources at various power output, we need new control and optimization 
mechanisms at both the transmission and distribution levels. The 
key control challenge is to match power-supply and -demand within  
acceptable voltage and frequency regulations. It demands for 
innovative centralized or decentralized control strategies based on 
fine-grained information of measured electrical values (e.g., 
voltage, power, frequency) of the load and the source. The difference in 
control actions impose different reliability and latency 
requirements. However, the performance requirements for control tasks are 
relaxed as compared to fault protection; \SI{99.9}{\percent} 
reliability and \SI{100}{\milli\second} latency while time synchronization 
accuracy is also less critical.

\begin{figure}[!t] 	
	\centering 		
	\includegraphics[width=0.9\linewidth]{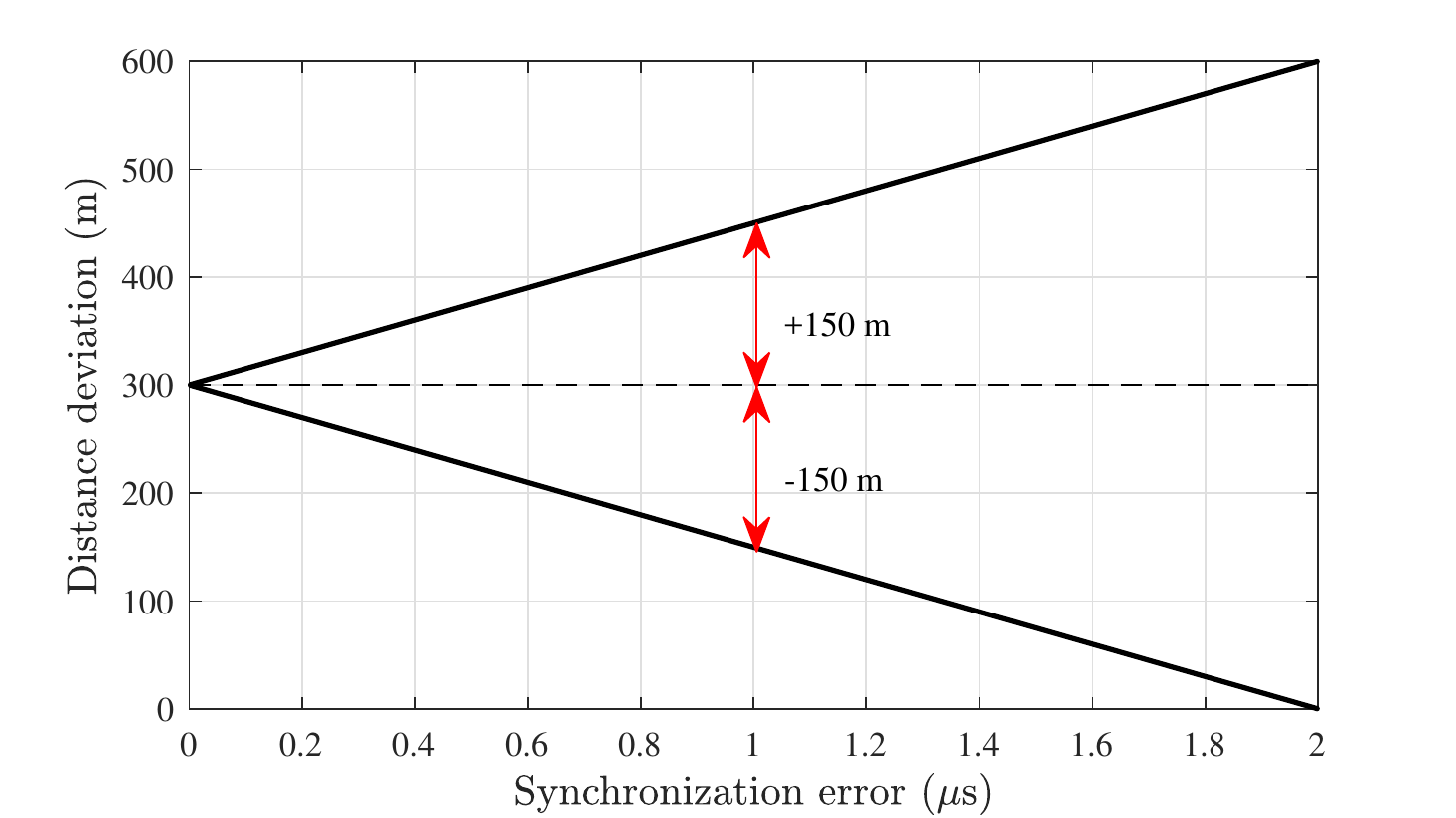}
	\vspace{-5pt}
 	\caption{Fault localization uncertainty w.r.t. synchronization error between two PMUs 
assuming the fault occurs at the middle of a \SI{600}{\meter} electricity line.} 	
	\label{fig:FaultLocation}
	\vspace{-10pt}
\end{figure}

\fakepar{Monitoring and diagnostics} Wide-area monitoring and diagnostics 
requires ultra-tight synchronization. For 
instance, the phasor measurement units (PMU) in electricity distribution systems, 
deployed along the electricity line, are used for measuring the 
electrical value for fault detection. When a fault occurs, the fault 
location generates two electric waves towards 
the both ends. The two PMUs, acting as UEs, at both end of the electricity line 
detect the waves by the change in monitored parameters and record 
the time of detection. A monitoring system can estimate the fault location 
based on the reported time information given that the PMUs are perfectly
synchronized. 
In most cases, synchronization accuracy of less 
than \SI{+-1}{\micro\second} is required to keep the uncertainty in fault 
location below \SI{300}{\meter} (see Fig.~\ref{fig:FaultLocation}). While, the reliability and latency requirements are relaxed for the monitoring; \SI{\sim 99}{\percent} and 
\SIrange{500}{1000}{\milli\second}. 

\section{Challenges}

\subsection{Heterogeneity of Industrial Networks/Standards}

URLLC can satisfy the requirements of most of the process and factory automation applications. However, it 
can not replace the existing industrial wired (e.g., industrial field bus, 
Ethernet of plant automation (EPA), Powerlink, Profinet) and wireless (e.g., 
WirelessHART, ISA 100.11a, WIA-PA\footnote{Wireless Networks for Industrial 
Automation-Process Automation: targeting \SI{99.9}{\percent} reliability and 
\SI{10}{\milli\second} latency. 
}, WIA-FA\footnote{Wireless 
Networks for Industrial Automation-Factory 
Automation: fulfills \SI{99.99}{\percent} reliability and 
\SI{1}{\milli\second} latency. 
}) 
communication systems completely, at once or in a short time. Two main 
reasons behind this are: 1) updating the complete communication 
infrastructure is not cost-effective, 2) lack of trust and/or unsuitability of 
replacing the existing wired system in critical manufacturing cells with 
wireless systems. As a result, a coexistence between the traditional 
wired/wireless networks and URLLC is a likely scenario that would
give a heterogeneous character to industrial networks.

\fakepar{Synchronization requisites} Owing to different 
requirements of industrial applications and different precision, 
resolution and stability of clocks running on the field devices, there are 
various existing synchronization solutions \cite{IEEE_Std_1588_2008}.  For 
the current coexisting wired-wireless
networks, 802.1~AS-2011 
\cite{IEEE_Std_8021AS_2011} offers accurate time synchronization. If the devices from heterogeneous 
domains (wired, traditional wireless and URLLC) need to collaborate and 
complete the production, cross-domain synchronization among heterogeneous 
industrial networks will be essential. 

\fakepar{Relative vs absolute time synchronization} For real-time 
communication, each industrial control entity should maintain a local clock 
to execute received commands at deterministic time 
instants. It works well under relative synchronization of the nodes with the 
controller. However, when multiple devices from different domains must 
operate concurrently, an absolute time synchronization among the 
devices will be mandatory. Hence, it is crucial that the devices are 
synchronized to an absolute time reference (e.g., UTC time) to perform 
coordinated operations.

\fakepar{Synchronization cases} The need of absolute 
time base at UEs arises from following situations. 
\begin{itemize}[leftmargin=*]
	\item \textit{URLLC domain}: In a URLLC network, it is plausible that the 
devices connected to different BSs may require absolute time synchronization 
to execute an operation synchronously. Therefore, to establish and 
maintain synchronization among the devices, the BSs must also be synchronized.

	\item \textit{Heterogeneous domain}: The time-sensitive devices might also 
belong to heterogeneous domains. For instance, 
an old wired device might need to 
perform a task synchronously with a new wireless device. As a result, 
it is not sufficient for the devices to synchronize with 
their respective controllers.
\end{itemize}
Accordingly the network synchronization architecture in a heterogeneous setup, 
must support synchronization 
among devices associated to i) a single BS, ii) different BSs, and iii) 
different 
domains (e.g., URLLC domain and local domain comprising traditional 
industrial networks.) 

\subsection{Time Synchronization in Smart Grids}

In today's smart grids, wired communication is often used for the 
applications discussed in Sec.~\ref{subsec:SmartGrids}. The synchronization 
among the measurement and control units, within a substation, is often 
maintained using IEEE 1588 or similar protocols. These protocols distribute 
the reference time, often acquired from GPS clocks, to the 
devices. To replace the wired infrastructure with 5G URLLC, a 5G 
network-based synchronization solution is required where only BSs may need to 
acquire time from GPS clocks. A built-in timing service of the wireless 
infrastructure can be cost-effective and it can avoid GPS 
vulnerabilities (e.g., jamming, signal reception). However, the challenge is 
to establish synchronization at the device level, which thus far is 
available to the BS level with limited accuracy 
(see Sec.~\ref{subsec:LTEnotSuits}).

\subsection{GPS for Reference Time}
\label{subsec:GPS}
GPS can align bases stations to a common time reference, however, GPS signal 
is not 
always available in e.g., indoor deployments, places with high constructions 
around and bad weather. The additional cost for GPS-based 
solution is also of a concern. For instance, in indoor industrial settings, 
the GPS antenna must be installed outdoors to ensure proper signal reception. 
Moreover, a feeder cable from the antenna to the receiver is required, along 
with an amplifier if the feeder cable is too long. To 
avoid GPS based solution, 5G network need to be upgraded to distribute 
the reference time to BSs. In addition, 5G network can be considered 
stable, adaptive and scalable as compared to wired/GPS solutions. However, the 
challenge is to devise OTA synchronization 
mechanisms to synchronization UEs to an absolute time through BS such that 
the devices are synchronized with each other.  
 
\subsection{Limitations of OTA Synchronization Solutions in LTE}
\label{subsec:LTEnotSuits}

\bgroup
\def\arraystretch{1.2}
\begin{table}[!t]
\centering
 \caption[Caption for LOF]{Phase/Time Synchronization Requirements of 
Different Application Features in LTE and LTE-A \cite{SyncSmallCell}\footnotemark{}} 
\label{tab:Tab1}
\centering
\begin{tabular}{|l|l|l|}
\noalign{\hrule height 1.2pt}
\textbf{Application}  & \textbf{Phase/Time Sync.}    &\textbf{Note} \\
\noalign{\hrule height 1.2pt}
LTE-FDD               & N/A &  -\\
\hline
LTE-TDD               & \SI{+- 1.5}{\micro\second}          & cell radius $
\leq$ \SI{3}{\kilo\meter}  \\
											& \SI{+- 5}{\micro\second}            & cell radius $>$ 
\SI{3}{\kilo\meter}\\
\hline
LTE-MBMS       				& \SI{+- 5}{\micro\second} 	         & intercell time 
difference \\
\hline
LTE-Advanced   				& \SIrange{+-1.5}{+-5}{\micro\second} & e.g., eICIC, 
CoMP\footnotemark{} \\
\noalign{\hrule height 1.2pt}
\end{tabular}
\end {table}
\egroup

\addtocounter{footnote}{-2} 
\stepcounter{footnote}\footnotetext{LTE requirs frequency 
synchronization of \SI{+-50}{ppb} as in earlier network generations, while GSM, 
UMTS and W-CDMA do not require phase synchronization.}
\stepcounter{footnote}\footnotetext{Coordinated multipoint (CoMP) and 
enhanced intercell interference coordination (eICIC) are interference 
coordination mechanisms. CoMP includes: joint transmission/reception, 
coordinated beamforming, dynamic point selection, and dynamic point blanking. 
}

In LTE, the distribution of accurate timing reference is 
limited up to radio BS level for LTE-TDD to avoid interference among adjacent cells. Meanwhile, the small cell 
deployments and the benefits of radio coordination features therein is now 
increasing the demands for accurate synchronization of BSs. Examples of radio 
coordination features are enhanced intercell interference coordination 
(eICIC) and coordinated multipoint (CoMP). Table~\ref{tab:Tab1} summarize the 
synchronization needs in LTE.    

\fakepar{Radio interface based synchronization} The current solutions, 
avoiding GPS- or backhaul network-based solutions, are 
focused on radio interface based synchronization (RIBS)-based OTA
mechanisms. In particular, the first solution uses network listening of the reference signals 
from neighboring BSs~\cite{3gpp.36.872}. However, the accuracy target is $<=$ \SI{3}{\micro\second}
 and the impact of propagation delay on synchronization accuracy is not 
considered. As the new interference coordination features in small cells 
require high synchronization accuracy, new RIBS-based solutions 
for intercell synchronization are under consideration in \cite{3gpp.36.898}, 
e.g.,  
\begin{itemize}[leftmargin=*]	
		\item Exchanging the reference signals (like IEEE 1588) between the neighboring base 
stations, which allow to calculate the propagation delay.

	\item Listening of reference signals from neighboring BSs by a target small 
cell, and compensating the propagation delay measured by a UE using timing 
advance (TA) while assuming negligible propagation delay between the UE and the target small 
cell. 
\end{itemize}

\fakepar{Implications for URLLC} The current activities in 3GPP are limited to 
intercell time synchronization while the tightest 
synchronization target among the 
base stations is \SI{+-0.5}{\micro\second}. The 
industrial automation and smart grid applications require the same order of 
jitter however among the collaborating UEs. If the UEs belong to the 
same BS, we need an OTA synchronization mechanism that can accurately 
synchronize UEs to the BS. However, achieving synchronization among UEs 
is challenging if collaborating UEs belong to different BSs. This is because 
the time alignment error between the base stations and the 
expected error in OTA synchronization procedure to distribute time to UEs will add up. 
Therefore, for URLLC applications, the synchronization among BSs
must be tight to achieve better synchronization at the device level.

\section{Enablers for OTA Synchronization} 
\label{sec:Enablers} 
In this section, we explore the existing signaling 
parameters in LTE that could be potential enablers of achieving 
ultra-tight synchronization for URLLC. Also, we discuss how these features 
could lead to a new synchronization architecture. 

\subsection{3GPP and OTA Time Synchronization}
%

\fakepar{TA Enhancement} Timing advance (TA) advances or retards the uplink 
transmissions of UEs in time relative to the distance-dependent propagation 
delay from the serving BS. A BS estimates the UE-BS propagation 
delay and issues TA updates to the UE in order to ensure that the 
uplink transmissions of all UEs are synchronized. In this way, 
the uplink collisions due to changing propagation delays are mitigated. 

TA negotiation occurs during network access and connected states. 
At network access, TA value is estimated at BS from 
the \textit{network access request} sent by UE.  If the request is successful, 
the BS sends a TA command in \textit{random access response} with \SI{11}{bit} 
value where $\textrm{TA} \in \{0,1,\cdots, 1282\}$. The TA command directs 
the UE to transmit its uplink frame by multiples of $16T_s$ seconds, i.e., 
$\textrm{TA} \times 16T_s$, before the start of the corresponding downlink 
frame, where $T_s$ is the 
sampling period. In connected state, TA is negotiated with periodic MAC control messages--to 
maintain BS-UE connection--which adjust the uplink timing of a UE relative to 
its current timing.
It is \SI{6}{bit} value 
i.e., $\textrm{TA} \in \{0,1,\cdots, 63\}$ while each command adjusts the 
UE's current uplink timing by $(\textrm{TA}-31) \times 16T_s$ seconds. The 
frequency of TA command is configurable as \{500, 750, 
1280, 1920, 2560, 5120, 10240\}, which corresponds to the maximum number of 
subframes sent in between two TAs. As the subframes are consecutive and each 
subframe is \SI{1}{\milli\second}, the timer duration can be interpreted as 
the number of milliseconds and 
offers a trade-off between the accuracy of transmissions' alignment and the network 
load. 

Note that $T_s$ is the basic unit of time in LTE, which is equal to 
\SI{32.55}{\nano\second}. Due to discrete nature of TA (multiple of 
$16Ts$), the propagation delay adjustment at UE is subject to an error of
\SI{260}{\nano\second}--half of the TA step. In addition, the delay estimation error at the BS is also affected by the multipath propagation especially in harsh channel conditions. The error could be large if the wrong TA bin is selected due to random error. Therefore, TA enhancement is required to get better synchronization accuracy.




\fakepar{SIB16 Enhancement} System information (SI) is an essential aspect 
of LTE air interface,  which is transmitted by BS over broadcast control 
channel (BCCH) \cite{3gpp.36.331}. SI is comprised of a static and dynamic parts 
termed as master information block (MIB) and system information blocks (SIB), 
respectively. MIB contains frequently transmitted essential parameters
needed to acquire cell information such as system bandwidth, 
system frame number, and physical hybrid-ARQ indicator channel 
configuration. It is carried on BCH transport channel 
and transmitted by physical broadcast channel (PBCH) every 
\SI{40}{\milli\second}. All SIBs except SIB1 are scheduled dynamically. SIB1 contains information 
including; cell access restrictions, cell selection information and 
scheduling of other SIBS. Unlike MIB, SIBs are 
carried on DL-SH and transmitted on physical downlink shared channel 
(PDSCH). SIB1 configures the SI-window length and transmission periodicity 
of the SI messages. Although SIB1 is transmitted with a fixed schedule of 
\SI{80}{\milli\second}, the resource allocation of PDSCH carrying all 
SIBs is dynamic. The resource allocation of PDSCH transmissions is indicated 
by \textit{downlink control information} message, which is transmitted on 
physical downlink control channel (PDCCH).

In time-aware applications, a UE can get a common time reference (UTC and 
GPS) contained in SIB16. However, time 
information in SIB16 has limited granularity (i.e., $\sim$\SI{10}{\milli\second}). To enable 
high accuracy synchronization service for URLLC applications, the granularity 
of SIB16 needs to be enhanced to \SI{}{\micro\second} or \SI{}{\nano\second} 
level. 

\subsection{New Synchronization Architecture}
\begin{figure}[!t] 	
	\centering 		
	\includegraphics[width=0.94\linewidth]{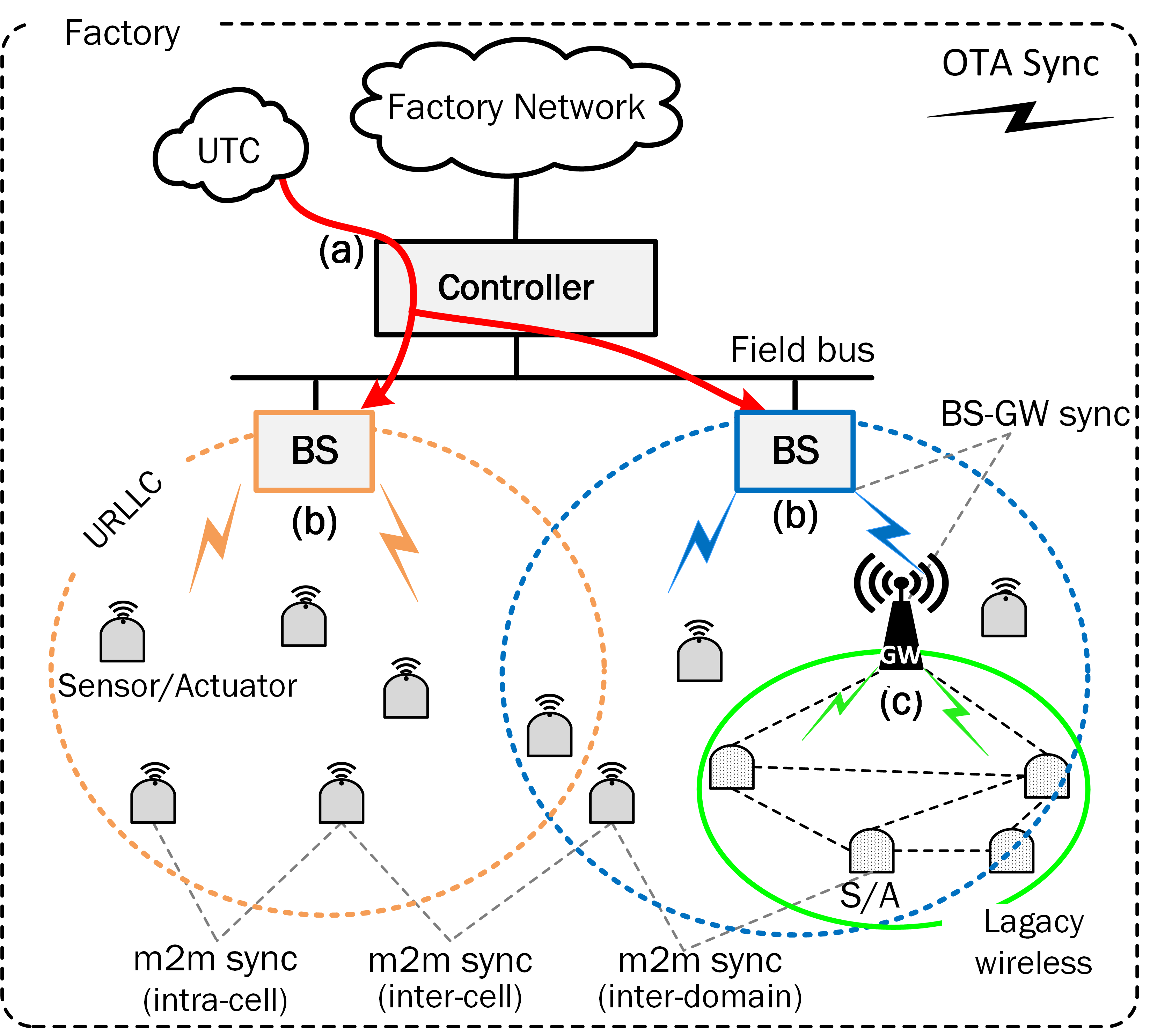}
	\vspace{-8pt}
 	\caption{Synchronization architecture to distribute time to obtain machine-to-machine (m2m) synchronization. Time flows as: (a) BSs synchronize with UTC time, (b) BSs distribute UTC time using OTA synchronization to URLLC devices and the legacy GW while the propagation delay is adjusted using TA, (c) the legacy GW acts as a master clock for its domain.}
	\vspace{-16pt}
	\label{fig:syncArch} 
\end{figure}

Considering emerging heterogeneity of industrial networks, a new ultra-tight 
synchronization architecture should provide: a) a flexible infrastructure for 
reference time, b) the absolute synchronization among devices. The 5G base 
stations can be used to provide time reference inside the URLLC domain and to 
the devices accessed via gateway. 
In this respect, each base station acts as a master 
clock while UEs/GWs as slave clocks. An example of such architecture is 
depicted in Fig.~\ref{fig:syncArch}, where the flow of time reference in an 
heterogeneous industrial network setting is as follows:
\begin{itemize}[leftmargin=*]
	\item Base stations acquire the reference time from a common source and act 
as 
master clocks for their associated devices. 
	\item GW/URLLC devices acquire the reference time from base stations using 
	OTA time synchronization procedure. While the GW acts as a master 
clock to its local domain and distributes time to the devices.
\end{itemize}
Accordingly, with already discussed signaling options in LTE, a few OTA synchronization enablers for URLLC are:


\fakepar{TA + SIB16} By mitigating the UEs-to-BS propagation delay based on TA, 
UEs can only establish a relative synchronization with the BS. 
Moreover, the 
indication of UTC time to the devices via SIB16 will suffer a loss in 
synchronization accuracy due to distance-dependent propagation delay. Therefore, the 
distribution of high-granularity UTC time with TA-based time offset 
adjustment can be a possible enabler to synchronize UEs to an 
absolute time reference. 

However, multiple factors can still disrupt the synchronization accuracy: for instance,  
1) the dynamic resource allocation for the transmission of SIB16 will add an 
uncertainty in UE-to-BS time-offset. To tackle it, either the timestamping of 
SIB16 must take place just before its transmission or the delay in the resource 
allocation must be statistically characterized and handled, 2) TA is only an 
approximation of propagation time. The statistical errors in 
TA must be estimated and analyzed under the impact of industrial 
wireless channels.

\fakepar{RIBS for UEs} Inspired by the RIBS based synchronization of small cells, a UE-BS synchronization scheme can be designed based on the exchange of uplink/downlink timestamped reference signals.
In comparison with TA+SIB16 scheme, this solution 
allows to calculate the effect of propagation delay in UE-BS 
time offset. Note that the reference signals could be the existing ones, however, must not 
conflict with the reference signals used in RIBS.


\fakepar{Dedicated RRC signaling} A bi-directional exchange of timing 
information between the UEs and the BS, as in IEEE 1588, can be another 
enabling solution to obtain time synchronization at the UEs. If properly 
exercised, the reference clock at the BS can be accurately distributed to the devices 
without an additional propagation delay estimator. 
To achieve this, the timing 
information can be exchanged via dedicated radio resource 
control (RRC) signaling. 
Considering the security concerns in industrial automation, the RRC signaling 
with integrity protection will also ensure that the timing information is 
reliable and not altered with fake time by an adversary.

However, the exchange of time information over dynamically scheduled 
RRC messages can directly affect the accuracy of time distribution. 
Therefore, the timestaming procedure of RRC messages must be 
scrutinized and/or these messages must be categorized as time critical to 
enable prioritized handling. Adversely, adding dedicated signaling for 
synchronization can result in an increase in the network load.

\section{Conclusions}
In summary, ultra-tight synchronization can be considered as the third axis 
of URLLC-model when targeting time-critical applications. Two important URLLC 
use cases, industrial automation and smart grids, demand accurate 
synchronization of the devices with an absolute reference time. As 5G URLLC 
will not replace the existing industrial bus systems 
completely, new interfaces with the existing wired/wireless systems are 
required.  The new interfaces must be designed carefully as an additional 
interface normally 
causes additional latency and jitter problems. Based on the existing signaling 
parameters in 5G radio interface and their enhancements, a certain level of 
device-level synchronization accuracy can be achieved. However, non-dedicated 
allocation of signaling resources can still lead to a time uncertainty that 
must be scrutinized. Therefore, a careful design of a 
synchronization service is required to avoid network congestion and to keep 
the device cost/complexity reasonable yet ensuring device-level synchronism.

\bibliographystyle{IEEEtran}
\bibliography{TSinURLLC}

\end{document}